# Ultra-Low Temperature Li/CF$_x$ Batteries Enabled by Fast-transport and Anion-pairing Liquefied Gas Electrolytes

Yijie Yin[1], John Holoubek[2], Alex Liu[2], Baharak Sayahpour[2], Ganesh Raghavendran[2], Guorui Cai[2], Bing Han[2], Matthew Mayer[2], Noah B. Schorr[3], Timothy N. Lambert[4], Katharine L. Harrison[5], Weikang Li[2*], Zheng Chen[1,2,6*], Y. Shirley Meng[1,2,6,7*]

[1]Materials Science and Engineering Program, University of California, San Diego, La Jolla, CA 92093, USA

[2]Department of NanoEngineering, University of California, San Diego, La Jolla, CA 92093, USA

[3]Department of Power Sources R&D, Sandia National Laboratories, Albuquerque, NM 87123, USA

[4]Department of Photovoltaics and Materials Technology, Sandia National Laboratories, Albuquerque, NM 87123, USA

[5]Nanoscale Sciences Department, Sandia National Laboratories, Albuquerque, NM 87123, USA

[6]Sustainable Power and Energy Center, University of California, San Diego, La Jolla, CA 92093

[7]Pritzker School of Molecular Engineering, University of Chicago, Chicago, IL 60637, USA

**Abstract:**

Lithium fluorinated-carbon (Li/CF$_x$) is one of the most promising chemistries for high-energy density primary energy storage system in applications where rechargeability is not required. Though Li/CF$_x$ demonstrates high energy density (>2100 Wh kg$^{-1}$) under ambient conditions, achieving such a high energy density when exposed to subzero temperatures remains a challenge, particularly under high current density. Here, we report a liquefied gas electrolyte with an anion-pair solvation structure based on dimethyl ether with a low melting point (−141 °C) and low viscosity (0.12 mPa×S, 20 °C), leading to high ionic conductivity (> 3.5 mS cm$^{-1}$) between −70 and 60 °C. Besides that, through systematic X-ray photoelectron spectroscopy integrated with transmission electron microscopy characterizations, we evaluate the interface of CF$_x$ for low-temperature performance. We conclude that the fast transport and anion-pairing solvation structure of the electrolyte brings about reduced charge transfer resistance at low temperatures, which resulted in significantly enhanced performance of Li/CF$_x$ cells (1690 Wh kg$^{-1}$, −60 °C; 1172 Wh kg$^{-1}$, −70 °C based on active materials). Utilizing 50 mg cm$^{-2}$ loading electrodes, the Li/CF$_x$ still displayed 1530 Wh kg$^{-1}$ at −60 °C. This work provides insights into the electrolyte design that may overcome the operational limits of batteries in extreme environments.

## Introduction:

Primary batteries serve an indispensable role in providing sustainable power in extreme environments which require long storage and operation life[1]. Thus, there is an escalating demand for primary batteries with high energy/power density and extreme-temperature adaptability[2]. Amongst the well-known primary batteries, Li/CF$_x$ presents itself as one of the most promising candidates for satisfying the above requirements[3]. At the same time, other chemistries, e.g., Li/Manganese oxide (Li/MnO$_2$), Li/Sulfur dioxide (Li/SO$_2$), and Li/Thionyl chloride (Li/SOCl$_2$), suffer from swelling[4], gas venting, and toxicity[5-6]. Li/CF$_x$ is a lightweight, safe, and highly stable system with a low self-discharge rate of < 0.5 % per year at room temperature with the highest theoretical energy density up to 2180 Wh kg$^{-1}$ (CF$_1$ based on active materials)[7]. However, the Li/CF$_x$ batteries suffer inferior rate and low-temperature (Low-T) performance due to the sluggish bulk electrolyte transport and increased charge transfer impedance[8-9]. To overcome the above challenges, the kinetic limitations of Li/CF$_x$ must be understood and addressed. These include (1) Li$^+$ diffusion through the solid electrolyte interface (SEI) and cathode electrolyte interface (CEI) layers[10]; (2) Li$^+$ solvation and de-solvation processes; (3) Li$^+$ diffusion through bulk electrolytes; (4) Li$^+$ insertion and/or diffusion in CF-CF layers[11]; (5) C-F bond breaking. Of the steps above, 1-4 are directly related to the electrolyte, indicating that the electrolyte plays a major role in governing the low-T behavior. However, current electrolyte research prioritizes the pursuit of performance rather than comprehensive understanding of the dominating factors governing low-T behavior.

Historically, electrolyte design for low-temperature Li/CF$_x$ batteries have prioritized low freezing point and low viscosity solvents to optimize the Li$^+$ transport. Tracing back to the effective conventional electrolytes for Low-T CF$_x$ batteries, NASA's Jet Propulsion Laboratory firstly reported an electrolyte formula consisting of 1 M lithium tetrafluoroborate (LiBF$_4$) coupled with 4:1 dimethoxyethane (DME): propylene carbonate (PC), which could deliver more than 600 mAh g$^{-1}$ capacity at C/40 rate under −40°C[12]. The optimized salt concentration and tris(2,2,2-trifluoroethyl) borate (TTFEB) additive further enhanced the specific capacity to around 300 mAh g$^{-1}$ at C/5 rate under −60 °C[9]. Additionally, the utilization of acetonitrile outperformed the DME system at both power capability (C/10) and low-temperature discharge performance (−60 °C)[13]. This was due to its improved ionic conductivity (5 mS cm$^{-1}$ to 11 mS cm$^{-1}$), facilitating bulk electrolyte transport at low temperatures. However, recent reports detailing the insertion of solvated Li$^+$ into the CF$_x$ lattice and the formation of a ternary intermediate C-(solvated Li$^+$-F) imply that the electrolyte solvation structure directly influences the charge-transfer resistance as well, which is known to be crucial at low-temperature[11],[14]. To this end, replacing strongly solvating DME with relatively weak solvating methyl butyrate (MB), which enabled an anion-pairing solvation structure, has been shown to improve both the high rate and low-temperature performance of Li/CF$_x$ cells. The authors demonstrated an improved rate performance (1 C, 834 mAh g$^{-1}$) and a 240 mAh g$^{-1}$ discharge capacity under −70 °C at 0.5 V cutoff voltage, although the formulated electrolyte delivered less than 1 mS cm$^{-1}$ ionic conductivity at −70 °C[15]. Therefore, the design criteria of low-T electrolytes for CF$_x$ batteries are either fast bulk ionic transport or formulating anion-pair solvation structure or integrating both parameters, where more recent studies demonstrated the anion-pair solvation structure may predominate the low-T discharge kinetics[16],[17]. However, the pursuit of both factors is mostly contradictory and rarely reported in the battery field. The formation of anion-pair structure requires the increase of salt concentration or the addition of inert diluents to form a locally high salt-to-solvent ratio, which reduces the ionic conductivity of the electrolyte and increases viscosity[18-19]. On the contrary, the dilute concentration electrolytes often offer the higher ionic conductivities, but they may suffer from the sluggish de-solvation process due to stronger Li$^+$-solvent coordination at reduced temperatures especially when using solvents with high solvating power[17, 20]. Apart from the above discussions, electrolytes also determine the properties of anode/electrolyte interphase (SEI) and cathode/electrolyte interface (CEI). For example, SEI formed on lithium metal vary at different temperature and is proven to affect the low-T lithium metal cycling efficiency[21]. Given the sensitivity of the CEI formed at $CF_x$ and the significant volume expansion after $CF_x$ discharge, there is no clear report on the chemical composition of the CEI at sub-zero temperature and its correlation with low-T performance.

Owing to the ultra-low melting point and viscosity of gaseous molecules[22], transformative liquefied gas electrolytes (LGE) based on hydrofluorocarbons (e.g. fluoromethane) were reported to deliver a superior electrochemical performance with Li/$CF_x$ at −40 °C although it offers < 1 mS cm$^{−1}$ ionic conductivity[23]. When paired with co-solvents, the formulated LGE improves the salt solubility and enables an anion-pairing solvation structure while maintaining a rapid transport at reduced temperature[24],[25]. These unique features of LGE strongly indicate a promising candidate for Low-T Li/$CF_x$ batteries.

Herein, we formulated a new LGE based on dimethyl ether ($Me_2O$) and PC, maintaining an ionic conductivity > 3.5 mS cm$^{−1}$ from −70 to 60 °C. Due to the weakly solvating power of $Me_2O$, the formulated electrolyte enables improved rate and low-temperature performance. The Li/$CF_x$ cell utilizing a 4.3 mg cm$^{-2}$ loading $CF_x$ cathode, delivered 780 mAh g$^{-1}$ (91 % room-temperature capacity retention) under 10 mA g$^{-1}$ at −60 °C. Moreover, when 50 mg cm$^{-2}$ $CF_x$ is utilized, the cell still displays 706 mAh g$^{-1}$ (84 % room-temperature capacity retention) at −60 °C and the average discharge voltage can be maintained above 2.1 V. Furthermore, a systematic study combining different advanced characterizations was conducted to figure out the improving mechanism, including both the bulk and interphase aspects.

## Results

An ideal electrolyte for ultra-low temperature and high-rate Li-CF$_x$ primary batteries should offer the lowest possible melting point (< −100 °C) and low viscosity. Besides, the electrolyte should easily de-solvate from its solvation shell, which brings about reduced charge transfer resistance[14]. The Me$_2$O shows an ultra-low melting point of −141 °C and a viscosity of 0.12 mPa×S at 20 °C, which outperforms DME with −58 °C and 0.46 mPa×S, acetonitrile (ACN) with −45 °C and 0.343 mPa×S, tetrahydrofuran (THF) with −108 °C and 0.456 mPa×S, and the recently reported methyl butyrate (MB) with −95 °C and 0.526 mPa×S **(Figure 1a)**. Among gaseous solvents, Me$_2$O endows higher salt solubility than fluoromethane (FM) and difluoromethane (DFM) owing to the higher Lewis basicity of the C-O-C than C-F[26], further enhancing electrolyte's ionic conductivity. In addition, Me$_2$O has been proved to offer excellent lithium metal compatibility at a wide temperature range[27]. Considering the above features, Me$_2$O is introduced to replace DME in the conventional LiBF$_4$-DME-PC formulations. We first optimized the ratio between Me$_2$O and PC to maximize transport properties and discharge performance. As shown in **Figure S1**, when the volume ratio reaches 6.5:1, the optimized electrolyte delivered the highest ionic conductivity of 3.54 mS cm$^{-1}$ at −70 °C and the highest room-temperature discharge capacity and nominal voltage. Furthermore, different lithium salts in 6.5:1 volume ratio of Me$_2$O:PC electrolytes have been evaluated, and we found LiBF$_4$ exhibited optimal CF$_x$ capacity utilization and discharge overpotential over lithium bis(fluorosulfonyl)imide (LiFSI) and lithium bis(trifluoromethanesulfonyl)imide (LiTFSI) salts at room temperature **(Figure S2)**, which is in alignment with previously reported results that LiBF$_4$ could reduce the activation energy for the charge transfer process[28]. Thus, the 1 M LiBF$_4$ in Me$_2$O: PC at a 6.5:1 volume ratio was formulated as the optimized electrolyte, hereby denoted as 1 M LiBF$_4$-Me$_2$O-PC. 1 M LiBF$_4$ in DME: PC with 6.5:1 volume ratio (denoted as 1 M LiBF$_4$-DME-PC), 1 M LiBF$_4$ in DME (denoted as 1 M LiBF$_4$-DME) and 1 M LiBF$_4$ in Me$_2$O (denoted as 1 M LiBF$_4$-Me$_2$O) are chosen as control systems for the mechanism study.

The ionic conductivities were measured to investigate the transport properties, as shown in **Figure 1b**. Owing to the superior physical properties of Me$_2$O, the 1 M LiBF$_4$-Me$_2$O-PC and 1 M LiBF$_4$-Me$_2$O demonstrate stable ionic conductivity from −70 to +60 °C. Among them, 1 M LiBF$_4$-Me$_2$O-PC invariably displayed > 3.5 mS cm$^{-1}$, higher than the electrolyte without PC. In contrast, although the conventional 1 M LiBF$_4$-DME-PC exhibits an ionic conductivity of > 4 mS cm$^{-1}$ before −10 °C, a large drop is observed (< 1 mS cm$^{-1}$) below −20 °C, which is due to the salt precipitation from the electrolyte (**Figure S3**). Similarly, severe ionic conductivity drops were observed for the other liquid 1 M LiBF$_4$-PC and 1 M LiBF$_4$-DME systems at reduced temperatures, mainly caused by the salt precipitation or the freezing of the electrolytes.

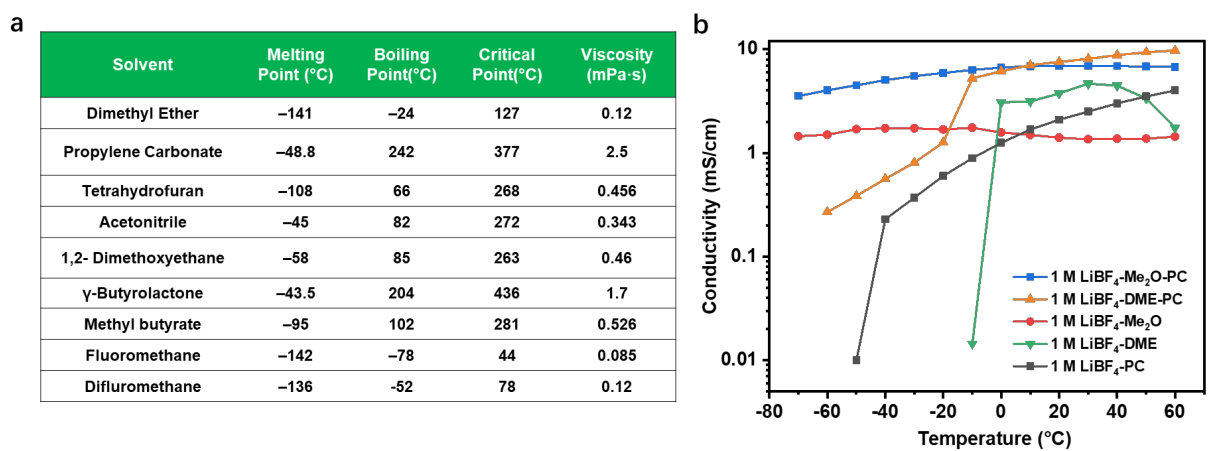

**Figure 1.** Design of the Low-T Electrolytes (a) Summary of physical properties of different solvents, data extracted from published works[29-30] (b) Measured ionic conductivity of the investigated electrolytes at different temperatures.

The solvation structure of the electrolyte influences the Li$^+$ de-solvation process[16], as commonly depicted by Molecular dynamics (MD) simulation and Raman spectroscopy[31]. Here, both techniques were applied to understand the effect of solvent selection on anion-pairing. 1 M LiBF$_4$-Me$_2$O-PC, 1 M LiBF$_4$-DME-PC, 1 M LiBF$_4$-PC, and 1 M LiBF$_4$-Me$_2$O were directly compared with the individual solvents and salt. Based on the Raman spectra in **Figure 2a**, the solvated BF$_4^-$ (B-F stretching) in Me$_2$O-PC solvents exhibits a blue shift compared with the DME-PC system, indicating more anions participate in the solvation shell[32]. As for the C-O-C stretching of Me$_2$O (**Figure 2b**), a slight peak broadening is observed in 1 M LiBF$_4$-Me$_2$O-PC and 1 M LiBF$_4$-Me$_2$O compared with the pure solvent, demonstrating the involvement of Me$_2$O in the Li$^+$ solvation shell[30]. As shown in **Figure 2c**, the bending mode of the PC ring also varies in different electrolytes. 1 M LiBF$_4$-PC and 1 M LiBF$_4$-Me$_2$O-PC showed an obvious additional solvated PC peak at 724 cm$^{-1}$ compared with pure PC and 1 M LiBF$_4$-DME-PC, where both spectra show the symmetric peak centered at 709 cm$^{-1}$. This suggests an increase of PC molecules participation in the solvation structure in the 1 M LiBF$_4$-Me$_2$O-PC and 1 M LiBF$_4$-PC electrolytes. Based on the above observations, the anion-pairing solvation structure of 1 M LiBF$_4$-Me$_2$O-PC is demonstrated in **Figure 2d,** which differs from the solvent coordinated solvation structure of 1 M LiBF$_4$-DME-PC.

MD simulations confirmed the observations from Raman spectroscopy. The simulation boxes contain 1 M LiBF$_4$-DME-PC (**Figure 2e**) and 1 M LiBF$_4$-Me$_2$O-PC (**Figure 2h**). After equilibration, the radial distribution functions (RDFs) for Li$^+$ in 1 M LiBF$_4$-DME-PC and 1 M LiBF$_4$-Me$_2$O-PC were computed, and the related results are shown in **Figure 2f** and **2i**. In terms of probability, it was found that DME predominates the solvation shell, whereas BF$_4^-$ anion and PC accounted for lower but comparable percentages (**Figure 2f-g**), resulting in an average Li coordination environment consisting of 2.3 DME, 0.39 PC, and 0.38 $BF_4^-$. On the other hand, the most probable coordinating species in 1 M $LiBF_4$-$Me_2O$-PC is $BF_4^-$, followed by PC and $Me_2O$ (**Figure 2i-j**), resulting in an average Li coordination environment consisting of 0.81 $Me_2O$, 1.1 PC and 2.4 $BF_4^-$. In addition, the DFT calculations suggest weaker binding between the $Me_2O$ molecule and $Li^+$ of −1.76 eV than the DME molecule and $Li^+$ of −2.84 eV (**Figure S4**), which is consistent with the Raman and MD results. Such anion-paired solvation structure has been proved to significantly benefit the $Li^+$ diffusivity (**Figure S5** and **Figure S6**) and de-solvation portion of charge transfer, resulting in facile kinetics and an improved low-temperature performance[33],[34],[14].

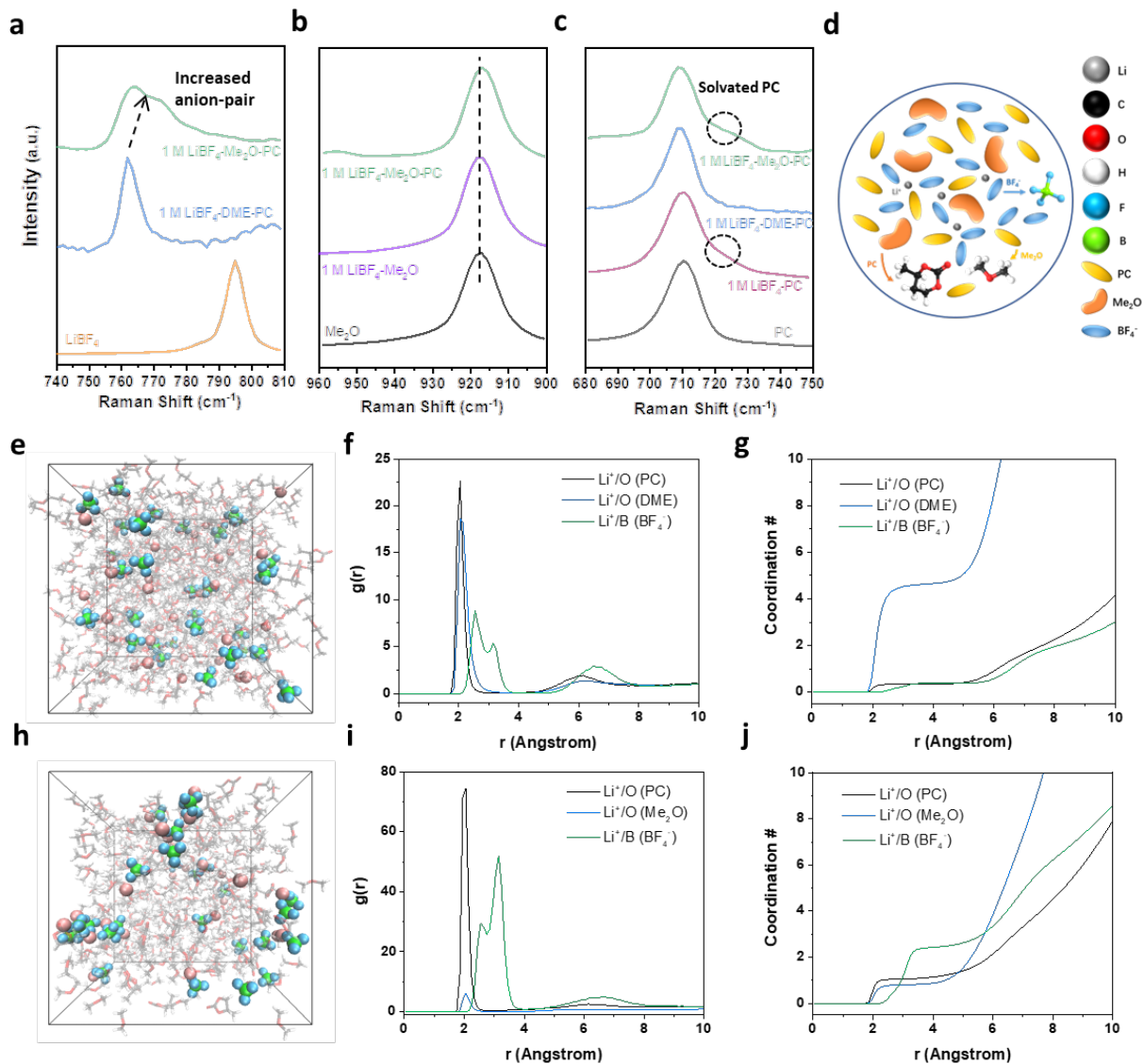

**Figure 2.** Raman spectra and simulated results of formulated and reference electrolytes. Raman spectra for (a) LiBF$_4$ salt in different solvents (B-F stretching), (b) Me$_2$O solvent in different electrolytes, and (c) PC solvent in different electrolytes. (d) Proposed solvation structure of formulated electrolyte. (e) Snapshots of the MD simulation cell containing 1 M LiBF$_4$-DME-PC. (f) (g) Li$^+$ radial distribution function and coordination number obtained from MD simulations of 1 M LiBF$_4$-DME-PC. (h) Snapshots of the MD simulation cell containing 1 M LiBF$_4$-Me$_2$O-PC. (i) (j) Li$^+$ radial distribution function and coordination number obtained from MD simulations of 1 M LiBF$_4$-Me$_2$O-PC.

Four operating temperatures (−70, −60, +23, +55 °C) were performed to evaluate the temperature-dependent discharge performance of Li/CF$_x$ cells in the formulated electrolytes. The discharge profiles of the cells with the 1 M LiBF$_4$-Me$_2$O-PC and 1 M LiBF$_4$-DME-PC electrolytes are shown in **Figure 3a** and **Figure 3b**, respectively. Under a current density of 10 mA g$^{-1}$, the two electrolytes delivered similar capacities at both 23 and 55 °C, comparable to the theoretical capacity (865 mAh g$^{-1}$) of the CF$_x$ (x=1). However, at reduced temperatures, the 1 M LiBF$_4$-Me$_2$O-PC electrolyte produced substantially improved performance than 1 M LiBF$_4$-DME-PC, providing 780 mAh g$^{-1}$ and 603 mAh g$^{-1}$ at −60 and −70 °C, respectively, with higher discharge voltage plateaus. In comparison, the 1 M LiBF$_4$-DME-PC electrolyte demonstrates reduced discharge capacities of 431 mAh g$^{-1}$ at −60 °C and 267 mAh g$^{-1}$ at −70 °C, respectively. This difference can be attributed to the higher ionic conductivities of the 1 M LiBF$_4$-Me$_2$O-PC electrolyte with higher Li$^+$ diffusivity and a facile de-solvation process enabled by anion-pair solvation structure, which further gives rise to the utilization of CF$_x$ at such low temperatures, as confirmed by the more prominent LiF peaks from X-ray diffraction of the discharged CF$_x$ (**Figure S7**). Interestingly, the cell employing 1 M LiBF$_4$-Me$_2$O delivered 708 mAh g$^{-1}$ capacity at −60 °C (**Figure S8**), which was lower than the cell using 1 M LiBF$_4$-Me$_2$O-PC, but outperformed both cells discharged in 1 M LiBF$_4$-DME and 1M LiBF$_4$-DME-PC, indicating Me$_2$O is more crucial than PC for the low-T performance.

To further evaluate the rate performance, Li/CF$_x$ cells were discharged at increased current densities of 1000 and 5000 mA g$^{-1}$ at room temperature. As shown in **Figure 3c**, the two electrolytes deliver similar capacities at a current density of 1000 mA g$^{-1}$. However, under 5000 mA g$^{-1}$, the 1 M LiBF$_4$-Me$_2$O-PC demonstrates a higher discharge capacity of 645 mAh g$^{-1}$ when compared to 603 mAh g$^{-1}$ in the 1 M LiBF$_4$-DME-PC. The electrolyte performance at reduced temperatures was also evaluated under increased current densities, as shown in **Figure 3d** for −60 °C and **Figure S9** for −70 °C. At −60 °C, the 1 M LiBF$_4$-Me$_2$O-PC retained 63.6% of the CF$_x$ theoretical capacity at a high current density of 300 mA g$^{-1}$ while the 1 M LiBF$_4$-DME-PC failed to discharge at 100 mA g$^{-1}$. At −70 °C, the 1 M LiBF$_4$-Me$_2$O-PC electrolyte again demonstrates improved performance against the reference electrolyte which failed to discharge at 100 mA g$^{-1}$. When using 50 mg cm$^{-2}$ CF$_x$ with 409 μm thickness (**Figure S10**), the 1 M LiBF$_4$-Me$_2$O-PC can discharge at 100 mA g$^{-1}$ with a higher voltage drop (down to 1.57 V) at room temperature (**Figure 3e**). When the cells are exposed to −60 °C, the 1 M LiBF$_4$-Me$_2$O-PC maintains 35.3 mAh cm$^{-2}$ capacity (706 mAh g$^{-1}$) at such extreme conditions (**Figure 3e**). By contrast, the 1 M LiBF$_4$-DME-PC delivered 855 mAh g$^{-1}$ capacity at room temperature but almost no capacity at −60 °C even with predischarge step (**Figure S11**). Even under 100 mA g$^{-1}$ current density at −60 °C, the cell using 1 M LiBF$_4$-Me$_2$O-PC still deliver 203 mAh g$^{-1}$ capacity with predischarge condition (**Figure S12**). In conclusion, the 1 M LiBF$_4$-Me$_2$O-PC enabled Li/CF$_x$ cells with high energy density at ultra-low temperatures when compared with other reported electrolytes, further reinforcing its promise to enable next-generation primary batteries in extreme environments (**Figure 3f, Supplementary Table 1**).

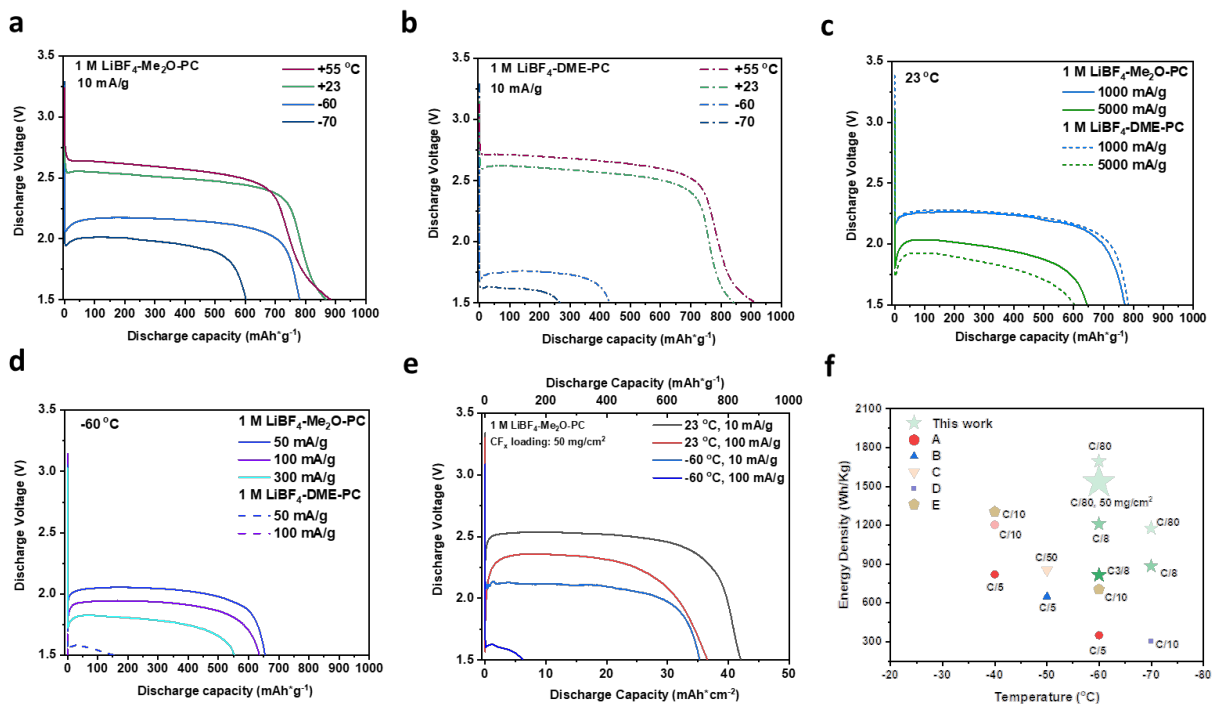

**Figure 3.** Electrochemical performance of $CF_x$ in different electrolytes (a) Measured electrochemical performance at a wide-temperature range of 1 M $LiBF_4$-$Me_2O$-PC. (b) Measured electrochemical performance at a wide-temperature range of 1 M $LiBF_4$-DME-PC. (c) Different current density discharge at room temperature. (d) Different current density discharge at −60 °C. (e) Different current density discharge at room temperature and −60 °C using high loading $CF_x$. (f) Summary of energy density at different temperatures from references (A[9],B[35],C[13], D[15], E[36]) and this work. The shades of color of each point indicates the current density and the size of each point describes the loading of the electrodes. The lowest reported loading is 1-2 mg cm$^{-2}$ and the highest one is 50 mg cm$^{-2}$. The 10 mA g$^{-1}$ current density used in this work roughly equals to C/80. It also applied to higher current densities where 100 mA g$^{-1}$ roughly equals to C/8 and 300 mA g$^{-1}$ roughly equals to C3/8.

To comprehend the outstanding performance delivered by 1 M $LiBF_4$-$Me_2O$-PC, we performed electrochemical impedance spectroscopy (EIS) to monitor the overall impedance during the different depths of discharge in both electrolytes. As shown in **Figure S13**, the EIS spectra are fitted following graphite/electrolyte interface model[37]. The bulk resistance ($R_b$) of solvated Li$^+$ in 1 M $LiBF_4$-$Me_2O$-PC remains stable over different depth of discharge states and is consistently lower than the 1 M $LiBF_4$-DME-PC, which aligns with the ionic conductivity results in **Figure 1**. In terms of the charge transfer impedance ($R_{ct}$) which represents the breakup of solvation shell of

Li$^+$, 1 M LiBF$_4$-Me$_2$O-PC has a R$_{ct}$ 2-4 times lower than that of 1 M LiBF$_4$-DME-PC before reaching the 20-hour discharge, where the turning points occur between the 10-hour and 20-hour discharge state. After the 20-hour discharge, the charge transfer resistance is significantly reduced in the 1 M LiBF$_4$-DME-PC but still higher than its counterpart. During the entire discharge, 1 M LiBF$_4$-Me$_2$O-PC possessed lower interfacial impedance (R$_{int}$), which indicates lower Li$^+$ diffusion barriers through the SEI/CEI. It is well-known that the interface plays an important role in the charge transfer kinetics, which is correlated to the de-solvation process of the electrolytes near the interface, the diffusion through CEI, and the chemistry and structure of CEI[38]. Considering the complexity in de-convoluting each step, X-ray photoelectron spectroscopy (XPS) was performed on the 10-hour discharged CF$_x$ at −60 ºC to investigate if the chemical composition of CEI determines the charge transfer impedance difference, and the data are shown in **Figure 4a-f**. Given that both samples were stopped at the same discharge capacity, the formed LiF and carbon should be the same in quantity. Based on the global survey of discharged CF$_x$, similar F, B, and O atomic concentrations can be observed over different etching times (**Figure 4b, c**). This indicates the similarity of interfacial chemistry in both electrolytes. We further examined the fine spectra of different elements. The C 1s from the pristine CF$_x$ electrode shows the characteristic structure of CF$_x$ materials, mainly containing C-C, C-F, and C-F$_2$ bonds (**Figure 4d**). After discharge, C-F/C-F$_2$ peaks decreased drastically, indicating the electrochemical reaction. Apart from that, CEI information can be depicted by O1s signal because the ether electrolytes are the source of extra oxygen. After 10-hour discharge, a new C=O appeared in both C 1s and O 1s spectra with a relatively weak intensity over different etching conditions, implying a thin CEI formed in both electrolytes. Interestingly, there is no obvious difference from both electrolytes in all XPS spectra, in addition to the more predominated C and F 1s signal (**Figure 4e, f**). When fully discharged to

1.5V, higher Li-F, less carbonyl group, and C-C signal can be observed in $CF_x$ discharged in 1 M $LiBF_4$-$Me_2O$-PC due to higher $CF_x$ utilization (**Figure S14)**. Based on the above analysis, we can conclude the CEI chemistry exerts nonobvious influence on low-T performance.

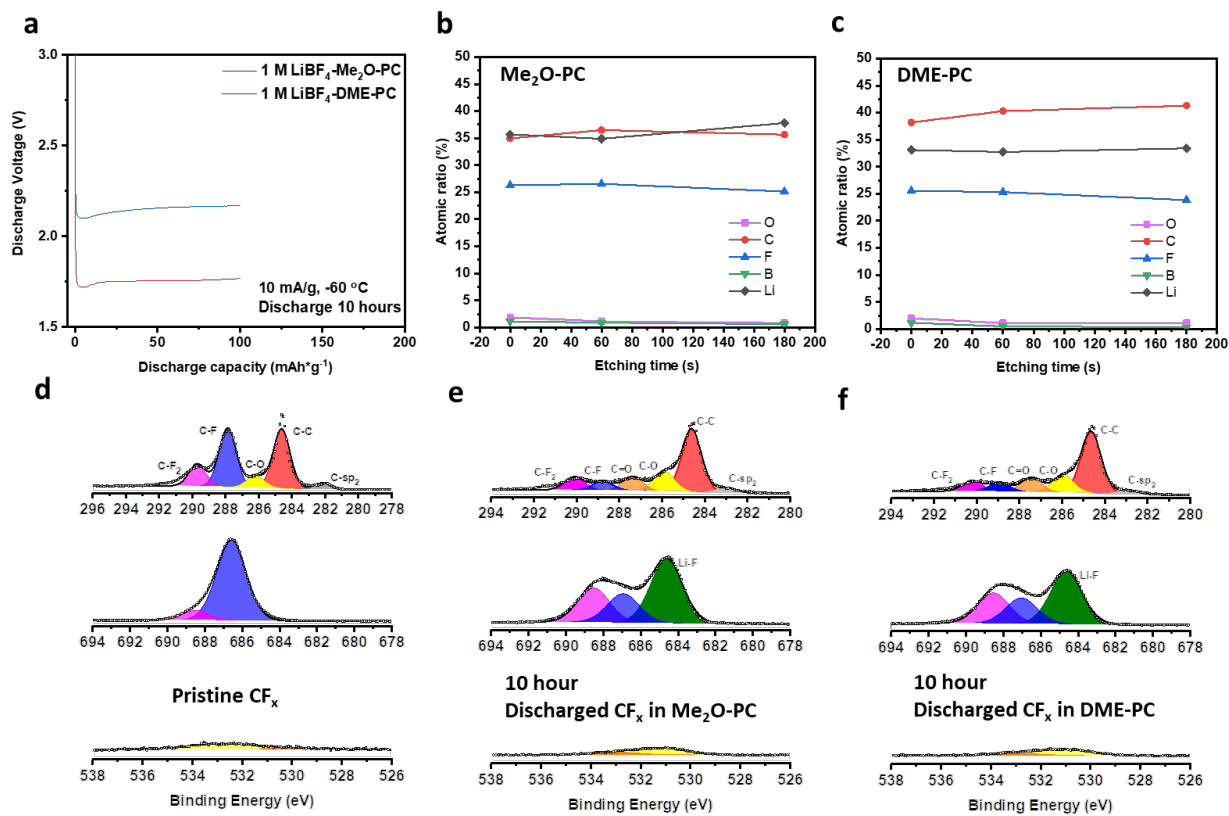

**Figure 4.** Global and local XPS analysis of the $CF_x$ at different states. (a) Voltage profiles of 10-hour discharged $CF_x$ in both electrolytes. Summary of atomic concentration of $CF_x$ discharged in 1 M $LiBF_4$-$Me_2O$-PC (b) and 1 M $LiBF_4$-DME-PC (c). (d) Local survey of pristine $CF_x$. (e) Local survey of 10-hour discharged $CF_x$ in 1 M $LiBF_4$-$Me_2O$-PC. (f) Local survey of 10-hour discharged $CF_x$ in 1 M $LiBF_4$-DME-PC. As for the XPS spectra, those represent C 1s spectra, F 1s spectra, and O 1s spectra from top to bottom view.

To understand the local $CF_x$ structure change during low-T discharge, scanning transmission electron microscopy-electron energy loss spectroscopy (STEM-EELS), high resolution transmission electron microscopy (HRTEM), and selected areal electron diffraction (SAED) were performed on $CF_x$ samples discharged at −60 °C in different electrolytes under 10 mA $g^{-1}$ (**Figure 5a-e, Figure S15-17**). Based on the STEM images and elemental mappings of discharged $CF_x$, a greater prevalence of Li was observed in 10-hour discharged $CF_x$ in the 1 M $LiBF_4$-$Me_2O$-PC compared to the 1 M $LiBF_4$-DME-PC at selected areas (**Figure 5a-b**). Both samples demonstrate the C and F elements with the new appearance of Li elements, where the Li distribution is more homogeneous in the discharged $CF_x$ in 1 M $LiBF_4$-$Me_2O$-PC. Coupled with EELS spectra (**Figure 5c-e**), both samples show Li-F feature as standard LiF sample, indicating the breaking of C-F bond and the formation of Li-F and graphitic carbon after 10-hour discharge, in consistent with our previous work. The inhomogeneity of LiF formation and scattered distribution of unreacted $CF_x$ from the $CF_x$ discharged in 1 M $LiBF_4$-DME-PC confirmed the sluggish transport / de-solvation properties of the 1 M $LiBF_4$-DME-PC electrolyte, which, in contrast, highlighted the superior performance enabled by the 1 M $LiBF_4$-$Me_2O$-PC with the homogeneous distribution of the discharged products. The fully discharged $CF_x$ were also evaluated, and the results were consistent with the observations from the 10-hour discharged samples (**Figure S16-17**). Considering the significantly reduced interfacial resistance obtained from the 1 M $LiBF_4$-$Me_2O$-PC electrolyte (**Figure S13**) for Li/$CF_x$ cell, the LGE should benefit the Li metal side as reported before[27], where $Me_2O$-based LGE demonstrates improved SEI structure compared with DME-based liquid electrolyte for lithium metal cycling at both room temperature and reduced temperature. Integrating with the above analysis, we can conclude that the structure of discharge products (LiF and graphitic carbon) appears similarly in both electrolytes and also places unimportant influences on low-T performance. Instead, bulk ionic transport and $Li^+$ de-solvation are more critical factors affecting the utilization of $CF_x$ and the distributions of discharge products.

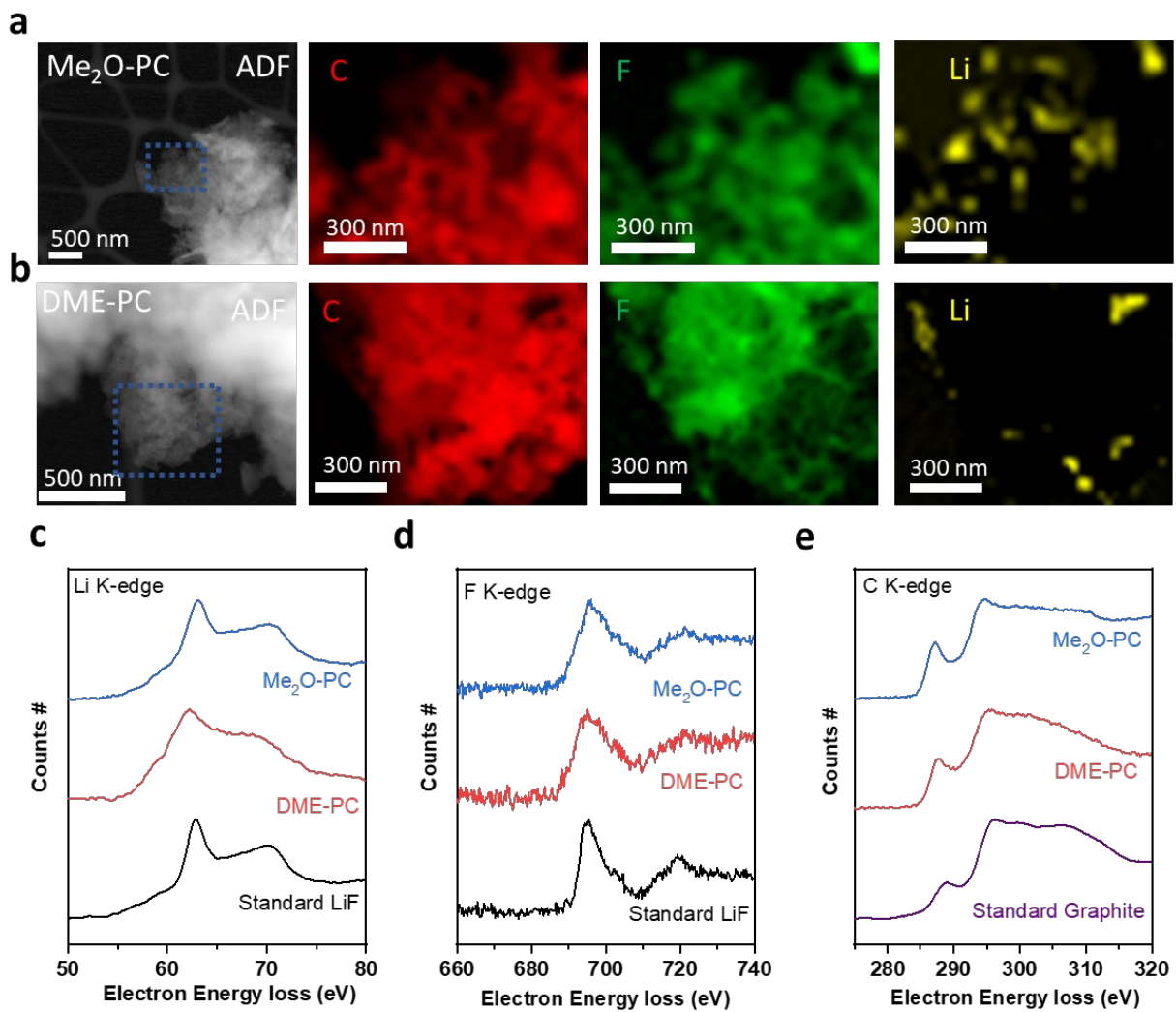

**Figure 5.** STEM-EELS, HRTEM and SAED of 10-hour discharged $CF_x$ at −60 °C. (a) STEM image and EELS mappings of discharged $CF_x$ in (a) 1 M $LiBF_4$-$Me_2O$-PC and (b) 1 M $LiBF_4$-DME-PC. EELS spectra of (c) Li K-edge, (d) F K-edge, and (e) C K-edge.

## Conclusions

In conclusion, 1 M $LiBF_4$-$Me_2O$-PC electrolyte has been well-formulated to improve the temperature-dependent and rate-dependent performance of Li/$CF_x$ primary battery. The optimized electrolyte demonstrated > 3.5 mS cm$^{-1}$ ionic conductivity through a wide temperature range of −70 to 60 °C. Raman, MD, and DFT simulations suggested the formulated electrolyte features an anion-pairing solvation of which the predominating Me$_2$O molecules have weak affinity with Li$^+$, facilitating the rate capability and low-temperature operation by affecting the de-solvation process while maintaining decent transport. Benefitting from the fast kinetics of the de-solvation and bulk transport, the optimized electrolyte enables high utilization of CF$_x$, demonstrating excellent rate performance at both room temperature and −60 °C and high energy over an extended operating temperature window (−70°C ~ +55°C). XPS and STEM-EELS revealed that the CEI chemistry had little impact on the low-T performance, highlighting the importance of electrolyte de-solvation and bulk transfer features. This work provides a route to enable high power and high energy density Li/CF$_x$ batteries operated in the extreme low-T environment, which may enlighten advanced primary battery designs with high energy and power in the future.

## Experimental Sections

**Materials**

Dimethyl ether (99%) was obtained from Sigma-Aldrich. The salts lithium bis(fluorosulfonyl)imide (99.9%) and lithium bis(trifluoromethane)sulfonimide (99.9%) were purchased from BASF and lithium tetrafluoroborate was purchased from Sigma-Aldrich. 1,2-dimethoxyethane (99.5%) and propylene carbonate were purchased from Sigma-Aldrich and stored over molecular sieves more than two days before formulating the electrolytes. The CF$_x$ powders were purchased from ACS material (GT1FS012). The CF$_x$ electrodes were made with an 8:1:1 ratio between active materials:PVDF:C65 and casted on Al foils. All casted electrodes were dried at 80 °C overnight before use. The CF$_x$ electrode loading is approximately 4.3 mg cm$^{-2}$.

Fabrication of 50 mg cm$^{-2}$ CF$_x$ cathodes was accomplished by forming and rolling a dough. First carbon black (Super-P) was mixed with a commercial carbon fluoride (Advanced Research Chemicals, ARC-5-R-175) in a 5:95 wt% ratio by using a mortar and pestle. Once thoroughly mixed, 5.6 wt% Teflon (60 wt% suspension in $H_2O$, Sigma-Aldrich) was added dropwise to the powder and mixing via mortar and pestle continued. With addition of binder the powder began to agglomerate, although not all powder adhered into one mass. To ensure a proper dough another 6.5 wt% of Teflon (wt% including previous Teflon addition) was mixed in with mortar and pestle. A small amount of isopropyl alcohol was used to wet the mixture and facilitate spread of Teflon among the carbon and CFx powders. Approximately 10 min of hand mixing after the second Teflon addition a dough formed that was free standing and did not shed powder. The dough was then rolled on a glass slab with a glass rolling pin to a thickness of ~0.5 mm and then dried at 80°C for 12 hr.

**Electrochemical measurements**

Ionic conductivity of different electrolytes was performed in custom fabricated pressurized stainless-steel cells with polished stainless-steel (SS 316L) as both electrodes. OAKTON standard conductivity solutions (0.447 to 80 mS $cm^{-1}$) were utilized to frequently calibrate the cell constant for the cells.

Electrochemical impedance spectroscopy was collected by a Biologic SAS (SP-200) system and the spectra were then fitted using ZView 4 software.

Battery discharging tests were performed using an Arbin battery test station (BT2043) from Arbin Instruments in custom designed pressurized stainless-steel cells. Li metal (FMC Lithium, 1 mm thickness, 3/8-inch diameter), separators and $CF_x$ electrodes were sandwiched, where Li metal serves as counter electrode and the $CF_x$ serves as working electrode. A three-layer 25μm porous PP/PE/PP membrane (Celgard 2325) was used for all the electrochemical tests. The electrolyte amount is flooded (> 50 g $Ah^{-1}$) for all electrolytes mentioned in this work.

For Li/$CF_x$ discharge tests in different temperatures, the cells were soaked at the testing temperature in a temperature chamber (Espec) for at least 2 hours before discharge. All room temperature discharge tests are performed without controlling the temperature. The pre-discharge of Li/$CF_x$ with 50 mg $cm^{-2}$ cathodes is performed at room temperature for 2-hour discharge using 10 mA $g^{-1}$.

**Material characterization**

The X-Ray Diffraction (XRD) measurements were done by a Bruker APEX II Ultra diffractometer with Mo Kα (λ = 0.71073 Å) radiations to check the crystal structures. The samples were prepared by scratching the cathode electrode and filling the capillary tubes inside an Ar-filled glovebox. All the cathode samples were not washed before these measurements.

Super-low-dose TEM/EELS techniques were developed for characterizing $CF_x$ structures. The discharged $CF_x$ cathodes were rinsed with DME to remove residual salt and dried at 80 °C under vacuum on a hotplate prior to analysis. The cathode powders were scratched from electrodes and put on a Cu TEM grid for all measurements. HRTEM samples were transferred into the TEM (ThermoFisher Talos 200X TEM operated at 200 kV), which is equipped with a CETA camera and low-dose system. The HRTEM images in panel D&F are acquired with an electron dose rate of ~200 e Å$^{-2}$ s$^{-1}$ for ~1s. The STEM (EELS Mapping) samples were also transferred into the ThermoFisher Talos 200X TEM. The TALOS microscope is equipped with a high-resolution Gatan imaging filter (Gatan Continuum 1069) for EELS mapping. The probe current utilized for EELS maps on the TALOS was approximately 140 pA.

Raman spectra of liquefied gas electrolytes were carried on Renishaw inVia confocal Raman microscope with an excitation wavelength of 532 nm. All spectra were calibrated with Si (520 nm) and analyzed by Wire 3.4 software developed by Renishaw Ltd. The Raman spectra measurements of $Me_2O$-based electrolytes were performed in a custom-built pressurized cell[31].

X-Ray photoelectron spectroscopy (XPS) was performed using a Kratos AXIS Supra DLD XPS with monochromatized Al Kα radiation (λ= 0.83 nm and hυ=1486.7 eV) under a base pressure <10$^{-8}$ Pa. To avoid moisture and air exposure, samples were transferred to the XPS chamber directly from a glovebox via air-tight transfer. All spectra were calibrated with hydrocarbon C-H C 1s (284.6 eV) and analyzed by CasaXPS software. To remove residual salt on the surface, all samples were rinsed with DME and dried in glovebox antechamber before analysis. The etching condition was set as an Ar1000+ cluster at 5 keV. The etching times were 60 s and 180 s.

**Computational Analysis**

Classical, fixed-charge Molecular Dynamics (MD) simulations were performed in LAMMPS using the General Amber forcefield for solvents and Li$^+$ with the anion described with the potentials of Doherty et al[39]. Liquid simulation boxes were constructed from random, amorphous distributions of the molecules, with compositions corresponding to the volume ratios and salt concentrations described above. In all cases the charges of the Li$^+$ and FSI$^-$ molecules were scaled to the optical dielectric of the solvents present in the system as employed by Park et al[40], which is 0.72 for DME/PC and 0.76 for Me$_2$O/PC. Periodic boundary conditions were applied in all directions.

For each system, the step size for all simulations was 1 fs. First, an initial energy minimization at 0 K (energy and force tolerances of $10^{-4}$) was performed, after which the system was slowly heated from 0 K to 298 K at constant volume over 0.01 ns using a Langevin thermostat, with a damping parameter of 100 ps. The system was then subjected to 5 cycles of quench-annealing dynamics in an effort to eliminate the existence of meta-stable solvation states, where the temperature was cycled between 298 K and 894 K at a ramp period 0.025 ns followed by 0.1 ns of dynamics at either temperature extreme with a total of 1.25 ns for all 5 cycles. After annealing, the system was equilibrated in the constant temperature, constant pressure (NpT ensemble) for 1.5 ns. The applied pressure was the 1 atm for DME/PC and 4.83 atm for Me$_2$O/PC, which was the experimental electrolyte pressure measured with Honeywell FP5000 pressure sensor at room temperature. The stresses in the system were isotropically resolved using the Andersen barostat at a pressure relaxation constant of 1 ps). Finally, we performed 10 ns of constant volume, constant temperature (NVT) production dynamics. Radial distribution functions and solvation snapshots sampled from the MD trajectory were obtained using the Visual Molecular Dynamics (VMD) software.

DFT binding energy calculations were performed using the Q-Chem 5.1 package. First, a geometry optimization step at the B3LYP//6-31+G(d,p) level of theory followed by single point energy calculations at the B3LYP//6-311++G** level of theory. Solvent binding energies were calculated as:

$$\Delta E = E_{Li^+ + solvent} - (E_{Li^+} + E_{Solvent})$$

## Author Information

### Corresponding Author

**Weikang Li -** *Department of Nano Engineering, University of California, San Diego, La Jolla, CA 92093*

**Zheng Chen -** *Materials Science and Engineering Program, Department of Nano Engineering, Sustainable Power and Energy Center, University of California, San Diego, La Jolla, CA 92093*

**Ying Shirley Meng** - *Materials Science and Engineering Program, Department of Nano Engineering, Sustainable Power and Energy Center, University of California, San Diego, La Jolla, CA 92093*
*Pritzker School of Molecular Engineering, University of Chicago, Chicago, IL 60637, USA*

### Authors

**Yijie Yin** - *Materials Science and Engineering Program, University of California, San Diego, La Jolla, CA 92093, USA*

**John Holoubek** - *Department of Nano Engineering, University of California, San Diego, La Jolla, CA 92093, USA*

**Alex Liu** - *Department of Nano Engineering, University of California, San Diego, La Jolla, CA 92093, USA*

**Baharak Sayahpour** - *Department of Nano Engineering, University of California, San Diego, La Jolla, CA 92093, USA*

**Ganesh Raghnavedran** - *Department of Nano Engineering, University of California, San Diego, La Jolla, CA 92093, USA*

**Guorui Cai** - *Department of Nano Engineering, University of California, San Diego, La Jolla, CA 92093, USA*

**Bing Han** - *Department of Nano Engineering, University of California, San Diego, La Jolla, CA 92093, USA*

**Matthew Mayer** - *Department of Nano Engineering, University of California, San Diego, La Jolla, CA 92093, USA*

**Noah B. Schorr** - *Department of Power Sources R&D, Sandia National Laboratories, Albuquerque, NM 87123, USA*

**Timothy N. Lambert** - *Department of Photovoltaics and Materials Technology, Sandia National Laboratories, Albuquerque, NM 87123, USA*

**Katharine L. Harrison** - *Nanoscale Sciences Department, Sandia National Laboratories, Albuquerque, NM 87123, USA*

## Data Availability:

All the data generated in this study are included in the Article and its Supplementary Information.

## Acknowledgements:

This work was supported partially by the Laboratory Directed Research and Development program (Project 218253) at Sandia National Laboratories, a multi-mission laboratory managed and operated by National Technology and Engineering Solutions of Sandia, LLC., a wholly owned subsidiary of Honeywell International, Inc., for the U.S. Department of Energy's National Nuclear Security Administration under contract DE-NA-0003525. The views expressed herein do not necessarily represent the views of the U.S. Department of Energy or the United States Government. This work was supported partially by an Early Career Faculty grant from NASA's Space Technology Research Grants Program (ECF 80NSSC18K1512) to Z.C. The SEM were performed in part at the San Diego Nanotechnology Infrastructure (SDNI) of UCSD, a member of the National Nanotechnology Coordinated Infrastructure, which was supported by the National Science Foundation (Grant ECCS-1542148).Y. Yin thanks Ich C. Tran for their help regarding XPS experiments performed at the University of California Irvine Materials Research Institute (IMRI) using instrumentation funded in part by the National Science Foundation Major Research Instrumentation Program (grant CHE-1338173). The authors acknowledge the use of Raman instrumentation supported by NSF through the UC San Diego Materials Research Science and Engineering Center (UCSD MRSEC), grant # DMR-2011924. The authors would like to acknowledge the UCSD Crystallography Facility.

## Author Contributions:

Y.Y., Z.C., and Y.S.M. conceived the original idea. Y.Y and W.L. designed the experimental plan. Y.Y., A.L. carried out the experiments. J. H. developed force field and conducted the MD simulations. G.R. and G.C. assisted with control experiments. B.S. performed the XRD characterization. Y.Y. and W.L. performed the XPS characterization and analysis. B.H. and Y.Y. performed the STEM-EELS characterization and analysis. N.B.S, T.N.L., and K.L.H. developed and fabricated high loading $CF_x$ electrodes. Y.Y., J.H., and W.L. prepared the manuscript with input from all co-authors. All authors have given approval to the final version of the manuscript.

## Declaration of Interests:

No potential competing interest was reported by the authors.